# POWER PARTICIPATION IN DIGITAL CITIZEN ENGAGEMENT IN SOUTH AFRICAN LOCAL GOVERNMENT: THE CASE OF MOBISAM

Caroline Khene, De Montfort University, caroline.khene@dmu.ac.uk

Ingrid Siebörger, Rhodes University, i.sieborger@ru.ac.za

Mamello Thinyane, United Nations University, mamello@unu.edu

Clement Simuja, Rhodes University, c.simuja@ru.ac.za

**Abstract:** A lack of service delivery and accountability are two characteristic challenges of numerous municipalities (local government) in South Africa. MobiSAM was introduced as a collaborative effort between a local university, civil society, residents, and local government to grow digital citizen engagement, facilitate two-way communication between local government and its citizenry, and contribute to the improved provision of basic services. Through the course of the project, it became clear that power, with respect to knowledge, was at play in the interactions or lack thereof between local government and citizens. This work-in-progress paper begins the journey of exploration of the power/knowledge dynamics at play in the MobiSAM project (using the SECI model and the Power Cube) by unpacking and understanding the nature of knowledge processed in the project and the associated knowledge creation processes that ensued between the different project stakeholders over time. The influence of power in determining the effective transfer of knowledge between key stakeholders of the project, for capacity building and organisational learning has emerged as an important issue in need of thorough investigation and critical analysis. This work in progress paper presents the preliminary framing of the research findings on the phases associated with the interplay of knowledge and power in the MobiSAM project, which are: 1) Realisation, 2) Navigating Responsiveness, and 3) Emergence.

## 1. INTRODUCTION

The concept of citizen engagement has at times been used as a façade to imply that effective strategies or approaches exist for two-way communication between government and citizens with respect to service delivery gaps and needs. Typically, citizen engagement is a state-led activity, providing spaces to solicit the views and involvement of citizens in decision-making around service delivery. Despite the values behind citizen engagement, the reality is far from its intended purpose; challenges experienced with local government in the Eastern Cape of South Africa are characterised by a lack of service delivery particularly in predominantly poor and rural communities (Algotsson et al., 2009; Ngumbela, 2021). Invariably, power is at play here. The dominant discourse around lack of service delivery links to the inequalities that have emanated from the past Apartheid legacy. However, high levels of imbalance in access to resources, infrastructure and social services still continue to plague the Eastern Cape Province and many similar vulnerable regions (Ferreira, 2021; Matolino & Kwindingwi, 2013; Nnadozie, 2013).

In an effort to address the power and information asymmetries that hamper service delivery and the right thereto, the MobiSAM project was introduced in 2012 as a digital citizen engagement initiative (Thinyane, 2013). MobiSAM stands for Mobile Social Accountability Monitoring and was introduced as a collaborative effort between a local university, civil society, residents, and a municipality (local government). The initiative exists as more than a digital application for residents to report service delivery faults but has grown to holistically influence communication ecologies around service delivery, and enable or incentivise stakeholder engagement that originally did not exist due to friction and suspicion among government and citizen activists or civil society (Machiri & Pade-Khene, 2020; Thinyane et al., 2017). In the process of implementing and operating MobiSAM (a pragmatist interpretivism and participatory action research project) from 2012 to the present 2021, researchers noted the power of knowledge and control in framing the direction of the project. The aim of the project was to contribute new knowledge to research and theory around digital citizen engagement in this context. What is particularly interesting in the current study, is the realisation by researchers (who are also local residents) of their ability to negotiate power and empower citizens (and even local government) to effectively communicate and articulate the existing challenges with respect to service delivery. Using the Power Cube, this research study considers the different aspects of power proposed by Gaventa (2006) (i.e., form, space, and level) in understanding the nature of power and power relations that drove the evolution of the MobiSAM project to what it is today. Given knowledge is the power at play (whether in reference to citizens' rights and awareness, or effective ways of enabling two-way communication between citizens and government in resource-constrained contexts), the paper also considers the SECI model of knowledge creation dimensions by Nonaka & Toyama (2003) to understand how knowledge as power is negotiated and socialised in the MobiSAM project over time. The study is therefore guided by the following research question:

> *What forms of knowledge and power frame the integration of a digital citizen engagement initiative over time in a resource-constrained local government context?*

The research study applied a pragmatist interpretive approach, facilitated by participatory action research. Pragmatist interpretivism is described by Ansell (2015, p. 13) as an approach "trying to understand how people (and social scientists) draw inferences in specific social contexts about the kind of situations they are in and about the intentions and motivations of others". Originating from political science, what differentiates this approach from pure interpretivism, is that the researcher is involved in the actual political or collectivist/advocacy process, and has to interpret and reflect on his/her experiences and that of his/her compatriots. Research around MobiSAM and its underlying concept of digital citizen engagement required researchers to engage in the iterative development and implementation, and reflection of actions and practices associated with the project. Here, the logical connection between theory and empirical data are developed in an iterative fashion, symbolising abductive reasoning, with a back-and-forth interaction between beliefs and action (Klein & Myers, 1999; Morgan, 2014; Walsham & Sahay, 2006). Even though this study is contextualised to the municipality that the MobiSAM project was implemented in, many local governments in South Africa, Africa, and other Global South contexts can learn from these experiences to consider strategies for resilient or effective digital citizen engagement integration.

## 2. DIGITAL CITIZEN ENGAGEMENT IN SOUTH AFRICA

The South African Constitution (Republic of South Africa, 1996) makes municipalities responsible for delivering basic services such as electricity, water, sanitation, and refuse removal. South African municipalities are responsible for ensuring service delivery, accountability, and participation at the local government level to realize a developmental democracy. Local government in South Africa has often fallen short of this mandate resulting in an inability of municipalities to deliver on basic services (Brand, 2018). Commentators have attributed this lack of delivery of basic services to a lack of accountability at local government level, particularly with respect to a lack of public participation (Heller, 2009; Kariuki & Reddy, 2017; Ndevu & Muller, 2018). Pravin Gordhan, the 2014 Minister for Co-operative Governance & Traditional Affairs, noted that the main problems faced by local government in South Africa are "a communication breakdown between councils and citizens; no accountability; political interference in administrations; corruption; fraud; bad management; violent service delivery protests; factionalism; and depleted municipal capacity" (Lund, 2014, p. 1). South Africa's Auditor General, in 2018, found that only 13% of South African local municipalities were fully financially compliant. This was attributed to a range of factors, namely, a lack of financial and management skills, political interference, infighting in councils, staffing shortages in key positions, and a lack of accountability (Brand, 2018; Kroukamp & Cloete, 2018).

Despite the numerous challenges (Brand, 2018; Kroukamp & Cloete, 2018; Lund, 2014), local government structures offer opportunities for increased citizen engagement. The immediacy of people's needs and the proximity of government to those who elected them provide increased motivation for participation. However, success is unlikely without meaningful, informed, and effective participation of citizens in government processes, and the provision of mechanisms and skills to hold service providers to account for their performance in managing public resources and delivering services (Heller, 2009; Kariuki & Reddy, 2017; Ndevu & Muller, 2018).

Social accountability can be viewed as the interaction of five elements to achieve its objectives (Helene Grandvoinnet et al., 2015), namely: state action, citizen action, information, citizen-state interface, and civic mobilization. The successful interaction of these elements is dependent on enablers that allow for the effective exchange of information between government and citizens. Two-way communication is fundamental, where citizens can either supply information or provide feedback to the state and vice versa (J. Gaventa & Barrett, 2012; Gigler & Bailur, 2014; H. Grandvoinnet et al., 2015). When information and communication technologies are used to enable this exchange through evidence-based citizen engagement and social accountability, this is referred to as digital citizen engagement (DCE) (World Bank, 2016, p. 16). The role of DCE and social accountability monitoring is to bring individual and collective voices together, to collaborate effectively through evidence gathered and shared (Peixoto & Fox, 2016). Traditional means of citizen engagement have included community scorecards and questionnaire surveys, to identify gaps in access to government services. However, the timely nature needed to gather real-time data that can be used effectively to address faults with respect to basic services, especially access to clean water (which has become even more critical in this time of the COVID-19 pandemic) are fundamental to dealing with the urgency of access to basic services. This calls for the need for DCE and social accountability monitoring to gather real-time evidence, around which citizens and government can engage as they work towards a negotiated cohesion for service delivery. MobiSAM attempts to offer citizens a methodology to engage in evidence-based Social Accountability Monitoring (SAM) and a set of mobile, web-based tools to facilitate the meaningful participation of citizens in local governance processes. However, the integration of such methods remains a challenge for many civic tech initiatives, as power and knowledge drive effective participation.

Emerging literature suggests that actors in a social system may exercise power in multiple ways: 'power over', and 'power to', 'power with', and 'power within' (Acosta & Pettit, 2013; Avelino & Rotmans, 2009). The 'power over' refers to actions of actors where there is control over another, for example actor A having more power over actor B in terms of dependency for resources (Avelino & Rotmans, 2009; Hannus & Simola, 2010). 'Power to' which is similar to the notion of 'agency', refers to the ability of actors to use their capacities and resources; for instance actor A can mobilise more resources than actor B in terms of economic power (Avelino & Rotmans, 2009; Lukes, 2005). 'Power with' describes the capabilities that arise out of collective action and collective agency. 'Power within' references empowerment. The power-knowledge relation in this research study is adopted to understand the involvement of actors or stakeholders (citizens, civil society, government and researchers) in the MobiSAM project, capable of producing knowledge that is either practical or theoretical, and a matter of episteme that is essentially contextual. The emphasis in this project has been on the role power plays in social change. Analysing power/knowledge relations can be complex in some cases since the power/knowledge relations are hidden within daily practices (Foucault, 1980). Therefore, it is important to envisage how power/knowledge relations could be analysed.

## 3. THEORETICAL FRAMEWORK

The starting point of exploring the power/knowledge dynamics at play in the MobiSAM project is unpacking and understanding the nature of knowledge processed in the project and the associated knowledge creation processes that ensued between the different project stakeholders. The research study therefore uses the SECI model and Power Cube to explore MobiSAM dynamics over time.

While knowledge fundamentally exists in human minds, its creation, processing, and use happens within broad "communities of interaction" that are shaped by the context; be it a social, cultural, or organizational context (Nonaka, 1994, p. 15). Knowledge therefore has both the *individual* dimension as well as the *collective* dimension through which it is created, shared, and communicated (Alavi & Leidner, 2001). One of the prominent knowledge management models that captures the varied dynamics of knowledge creation is the SECI model. The SECI model of knowledge creation was originally proposed for managing knowledge creation processes within organizations, and was centered on the social interactions around tacit and explicit knowledge, through which new knowledge is created by individuals supported by elements within the organization that articulate and amplify the knowledge (Nonaka, 1994). The SECI model identifies 4 cyclical modes of knowledge creation that ensue between tacit and explicit knowledge:

1. **Socialization**: a tacit-to-tacit knowledge creation process, which occurs in contexts of shared experiences and practice between individuals, and is typically encapsulated in organizational culture.
2. **Externalisation**: a tacit-to-explicit knowledge creation process, which occurs when instruments such as policies, guidelines and best practices are utilized to formalize, codify, and document organizations' knowledge.
3. **Combination**: enables the explicit-to-explicit knowledge creation process through supplementation, combination, analysis, and categorization of existing knowledge.
4. **Internalisation**: the explicit-to-tacit knowledge creation process which typically occurs through organizational learning practices.

The SECI model has not been without criticism - that it has been formulated from a very specific organizational and cultural context (Haag et al., 2010; Li & Gao, 2003), and that it lacks empirical validation. However, it has been shown to provide an effective framework for unpacking the key elements of knowledge management processes (Natek & Zwilling, 2016; Nonaka & von Krogh, 2009); for studying knowledge creation processes in different domains (Faith & Seeam, 2018); and for understanding the role of technology in knowledge creation processes (Alavi & Leidner, 2001; Dávideková & Hvorecký, 2017). When one considers the complex interactions and knowledge flows between the different stakeholders in DCE interventions, the relevance of the SECI framework as an analytical instrument becomes immediately apparent. It helps to locate actants, both human and non-human, across the four knowledge creation processes, and also to explore the role of technology to support those processes. For example, information management and data analytics systems would tend to support processes framed around explicit knowledge (e.g. combination), while interactive communication and social media systems would better support tacit knowledge processing around sharing of experiences and insights between users.

The overarching aim of DCE should be around the effective distribution of knowledge to evoke service delivery change. Typically, power is closely aligned with who has the knowledge to manipulate or drive service delivery operations. In many local government contexts that are resource constrained, knowledge is a monopoly in the hands of government officials or even (affluent) citizens who stand to gain from service inequality (Gaventa & Cornwall, 2006). Nonetheless, social accountability monitoring, with the help of DCE aims to challenge the power of the dominant actor(s), as citizens (especially from marginalised areas) begin to develop a new awareness of their realities and their rights. Gaventa's (2006) Power Cube coupled with the SECI model helps to unpack the evolution and resilience experienced in the MobiSAM project.

The Power Cube framework was born out of citizens "efforts to claim political, economic and social rights" (Gaventa, 2006, p. 26). It facilitates the interrogation of power to understand how and if participation and the voice of the citizens can and does have influence to promote inclusion and social justice. Three forms of power (which can be interrelated) are described in relation to the forms, spaces, and levels of power in which citizen engagement takes place:

1. **Spaces**: Gaventa (2006, p. 26) describes 'spaces', quoting (Lefebvre, 1992, p. 24) as "a social product …a dynamic, humanly constructed means of control, and hence of domination, or power".  There are three levels of spaces: closed spaces, invited spaces and claimed/created spaces.  Spaces tend to be dynamic in their existence and in their relationships to one another; continually opening and closing through transformation, co-operation, resistance, and legitimacy. Whoever creates the space is likely to hold the power in that space.
2. **Levels**: refer to where the "critical, social, political and economic power resides" (Gaventa, 2006, p. 27).  These levels are defined as local, national, and global.  These levels of power are interrelated; local forms of power are "being shaped in relationship to global actors and forces, and in turn, local action affects and shapes global power" (Gaventa, 2006, p. 28).
3. **Forms** of power refer to the "dynamics of power that shape participation" and are defined as visible, hidden, and invisible (or internalised) forms of power (Gaventa, 2006, p. 26; Lukes, 2005).  These forms of power place boundaries on participation to exclude certain actors or viewpoints from engaging in arenas of participation.

Gaventa (2006, 2019) notes that real transformative change happens when power is addressed at each dimension simultaneously. This is not a straightforward process, as we demonstrate in the MobiSAM project, which is still ongoing, and challenged by contextual factors.

## 4.     POWER PARTICIPATION IN MOBISAM

The concept of knowledge and knowledge transfer (Szulanski, 1996) in MobiSAM is not new – a previous publication (Pade-Khene, 2018) focused on developing digital literacy around the creation and use of MobiSAM. However, what became apparent over time was the influence of power, in determining the effective transfer of knowledge between key stakeholders of the project, for capacity building and organisational learning. Here, knowledge refers to both civic literacy and responsiveness literacy (Pade-Khene, 2018). In the analysis, two knowledge perspectives with respect to these types of literacy frame the discussion: '*state of mind*'[1] and '*capacity*'[2] (Alavi & Leidner, 2001). Civic literacy considers that citizens possess the domain knowledge, ability, and capacity to make sense of their political world, and hence effectively act individually or collectively to hold government accountable and demand for key services (Henninger, 2017). Responsiveness literacy on the other hand considers the domain knowledge of local government to respond to and acknowledge citizen requests, and engage on communication platforms with citizens. One would consider responsiveness literacy as a choice, *rather* than an element of 'capacity' or 'state of mind' as a perspective of knowledge (Alavi & Leidner, 2001) – where local government exercises their political power in decision-making. However, understanding citizen service delivery issues, their rights, and the ability of citizens to exercise those rights produces an understanding of and respect for accountability of government performance. Figure 1 provides an illustrative diagram that represents how the authors have begun to reflect on and analyse the progressive developments in the MobiSAM project, based on the interplay of knowledge and power over time. The interplay of knowledge and power vary in each of the three phases (1. Realisation, 2. Navigating Responsiveness, and 3. Emergence) that MobiSAM has experienced (see Table 1). Future publication will focus on defining, discussing and further evidencing these phases, in relation to the mode of knowledge dominant in each phase, and the spaces, forms, and levels of power influencing the transfer of knowledge.

---

[1] a cognitive **state of mind** of knowing and understanding

[2] a **capability** and a potential to influence action, build competency and know-how.

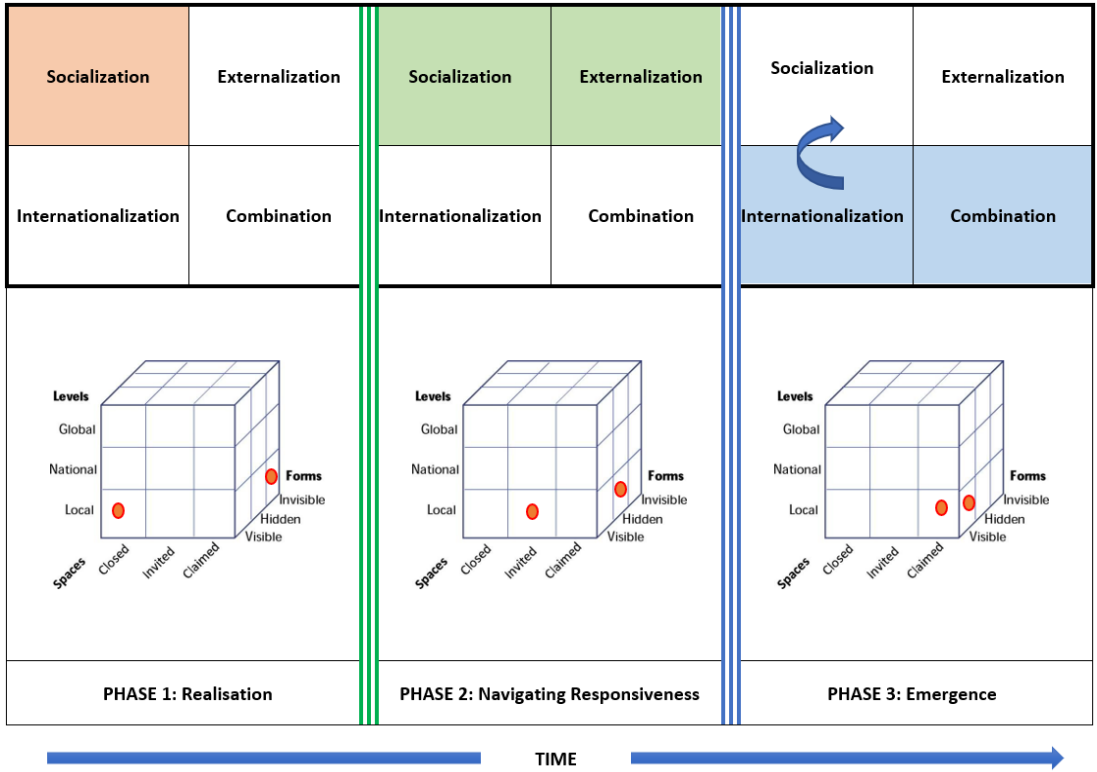

**Figure 1. The Phased Interplay of Power and Knowledge in the MobiSAM Project**

| Phase | Description |
|---|---|
| 1. **Realisation** | Associated with the introduction of MobiSAM, the realisation of the emanating challenges of service delivery and the lack of effective citizen engagement - the focus of the project was on empowering citizens, which the government tended to characterise as adversarial. Socialisation dominated as a mode of knowledge creation, as knowledge around service delivery issues tended to be shared among individuals - due to mistrust and despondency among citizens who complained among each other rather than officially report on an issue, as well as the power of government to withhold or dominantly share knowledge with each other. |
| 2. **Navigating Responsiveness** | This formed what was referred to as MobiSAM 2.0, with an aim to incentivise government (building responsiveness) and citizens to buy-into digital citizen engagement. A less adversarial approach was applied. The operation model incorporated strategy formulation with key stakeholder partners, iterative development of the application to account for needed requirements, citizen training, and evaluation to address uncertainty. The overall aim was to build civic literacy and responsiveness literacy - two key elements that enabled the use of MobiSAM to report and advocate for service delivery. As knowledge around service delivery became more explicit, externalisation as a mode of knowledge creation began to dominate. |
| 3. **Emergence** | MobiSAM has had to become organic and dynamic - its continued existence has been left to unfold, as researchers have completed the project but continue to support it with limited resources to ensure its continuity. The focus here is incentivising key actors such as civil society and government to support the continuity of MobiSAM, in however they see fit - designing it to support its own activities. For example, MobiSAM is seen as a vital source of communication by the government, and civil society sees it as a tool to generate evidence-based data to advocate for change. Other platform technologies such as WhatsApp and Facebook have been used in conjunction with the MobiSAM platform, to enable other forms of communication that are more accessible. The dominant mode of knowledge creation is Combination and Internalisation, as MobiSAM has become integrated into government's service delivery operations. |

**Table 1. Description of MobiSAM Phases**

## 5.  CONCLUSION

Challenges experienced with local government in the Eastern Cape of South Africa are characterised by a lack of service delivery and accountability particularly in predominantly poor and rural communities, despite the responsibilities thereof written into the Constitution of South Africa. MobiSAM was introduced as a collaborative effort between a local university, civil society, residents, and a municipality (local government) to grow digital citizen engagement, facilitate two-way communication between local government and its citizenry, and contribute to the improved provision of basic services.  Through the course of the project, it became clear that power, with respect to knowledge, was at play in the interactions or lack thereof between local government and citizens.  The power-knowledge relation in this research study is adopted to understand the involvement of actors or stakeholders (citizens, civil society, government, and researchers) in the MobiSAM project, capable of producing knowledge. The emphasis being on the role power plays in social change.  While the concept of knowledge and knowledge transfer is not new to MobiSAM, the influence of power in determining the effective transfer of knowledge between key stakeholders of the project, for capacity building and organisational learning has emerged as an important issue in need of thorough investigation and analysis.  Future research and analysis will focus on defining and discussing these phases, in relation to the mode of knowledge (SECI) dominant in each phase, and the spaces, forms, and levels of power (Power Cube) influencing the transfer of knowledge.